\documentclass[preprint,showpacs,showkeys,preprintnumbers,amsmath,amssymb]{revtex4-1}
\usepackage{graphicx}% Include figure files
\usepackage{dcolumn}% Align table columns on decimal point
\usepackage{bm}% bold math

\begin{document}
\input{epsf}
\preprint{APS/123-QED}

%\begin{document}
%\preprint{APS/123-QED}
%\title{ITERATION PROCEDURE FOR THE $N$ - DIMENSIONAL SYSTEM OF LINEAR EQUATIONS}

\title{Iteration Procedure for the $N$-Dimensional System of Linear Equations}

\author{Avas V. Khugaev\footnote{Permanent address: Institute of Nuclear Physics, 100214 Tashkent, Uzbekistan}}%
\email{avaskhugaev@yahoo.com}
\affiliation{Bogoliubov Laboratory of Theoretical Physics, \\
Joint Institute of Nuclear Research, 141980 Dubna, Russia}
% %%%%%%%%%%%%%
\author{Renat A. Sultanov}
\email{rasultanov@stcloudstate.edu}
\affiliation{Department of Information Systems and BCRL,
%Business Computing Research Laboratory, 
St. Cloud State University,
367B Centennial Hall, 720-4th Avenue South, St. Cloud, MN 56301-4498, USA}
%%\textbackslash\textbackslash
% %%%%%%%%%%%%%%%%%%
\author{Dennis Guster}
\email{dcguster@stcloudstate.edu}
\affiliation{Department of Information Systems and BCRL,
%Business Computing Research Laboratory, 
St. Cloud State University,
367C Centennial Hall, 720-4th Avenue South, St. Cloud, MN 56301-4498, USA
% with \\
}%

\date{\today}% It is always \today, today,

\begin{abstract}
A simple iteration methodology for the solution of a set of a linear algebraic equations
is presented. The explanation of this method is based on a pure geometrical interpretation
and pictorial representation. Convergence using this method is obtained and a simple numerical example is provided.
\end{abstract}


\maketitle

\section{Introduction}

In various scientific fields there is a need to reduce the initial problem to a set of linear algebraic equations.
This is true in physics when solving eigenstates or eigenfunction problems in quantum mechanics. Also, it is applicable
in applied physics and mathematics in the areas of linear programming, optimization methods and the root mean square
method. Therefore this reduction of the initial problem to a set of linear algebraic equations can be viewed as a critical step in
obtaining the correct solution to the initial problem. 

The methods used in obtaining a solution for a set of
the linear algebraic equations are well known. Typically, the process begins with Cramer's formulas,
deriving a final solution from the linear systems coefficients or similar
numerical methods, then using a Gaussian method to obtain further improvement and modifications or even applying the Jacobi
method to obtain further refinement.

However, the real problem starts with the attempt to effectively solve a
set of linear equations containing a large number of variables in which Cramer's method is not effective, because we need to calculate
determinants, which in itself is a problem when a large number of variables are involved because there will be a loss
of accuracy and the numerical stability of the calculation procedure is reduced. This would also apply in the application of
both the Gaussian and Jacobi methods as well. In obtaining practical solutions we may also have observed linear algebraic set
coefficients with error bars which makes the solution obtained even more sensitive to the
method used in obtaining a numerical solution. 

In practice one of the main requirements desired in the numerical
method selected is its stability and fast convergence in obtaining the final solution. In this paper
the goal is to present a simple and rather convenient blueprint to describe the programming iteration
method that could be used to solve the linear algebraic equation problem mentioned above.
\vspace{3mm}

\section{Problem Reformulation. Geometric picture}

In this section we want reformulate the algebraic problem involving the
linear system solution using a geometrical language. Therefore, we need to obtain a number such as
$x_k$, which satisfies the set of the linear system:
\begin{eqnarray}
A_{11}x_1+A_{12}x_2+...+A_{1n}x_n=B_1\nonumber \\
A_{21}x_1+A_{22}x_2+...+A_{2n}x_n=B_2\nonumber \\
\ldots\ldots\ldots\ldots\ldots\ldots\ldots\ldots\ldots\ldots\ldots\nonumber\\
A_{n1}x_1+A_{n2}x_2+...+A_{nn}x_n=B_n
\label{eq1}
\end{eqnarray}
which can be written in a more convenient way, as:
\begin{equation}
\sum^{n}_{k=1} A_{ik}x_{k}=B_{i}\,.
\label{eq2}
\end{equation}
where $A_{ik}$ and $B_{i}$ completely define the set of linear equations and
have the following bounds $1\leq i\leq N$. From a geometrical point of view we can consider every
line in the set of linear equations, for any fixed number $i$, as a hyperplane
in the $n$ - dimensional space of the solution set. This means, that we can consider
the solution as a vector, defined by $\vec X=(x_1, x_2,..., x_n)$, such that the
matrix $A_{ik}$ transforms the vector $\vec X$ into the vector $\vec B$ which can be defined
as $\vec B=(B_1, B_2,..., B_n)$. For instance, let's take some fixed point
$P\in (\alpha_i)$, where ${\alpha_i}$ is a $i$-th hyperplane, such that
$P(x^{p}_{1},x^{p}_{2},...,x^{p}_{n})\in (\alpha_i)$. Then for the arbitrary point
$Q\in(\alpha_i)$ which also belongs to this hyperplane, we can write, that the scalar
product $\vec{A_i}\cdot\vec{QP}=0$, where $\vec{A_i}$,  is defined as
$\vec{A_i}\equiv (A_{i1},A_{i2},...,A_{in})$ is $\perp$ to the $(\alpha_i)$ vector.
This case can be described as:
\begin{equation}
\sum^{n}_{k=1}A_{ik}(x_{k}-x^{p}_k)=0\longleftrightarrow
\sum^{n}_{k=1}A_{ik}x_{k}=\sum^{n}_{k=1}A_{ik}x^{p}_k=B_{i}\,.
\label{eq3}
\end{equation}
Therefore, we can see, that from a geometrical point of view the solution of
a linear set of algebraic equations is some point in $n$ space, which is an intersection of all
$n$ hyperplanes. One of the possibly solutions therefore, is the construction of an
iterative process, which will allow convergence at this point.

\section{Normalization procedure}
\setcounter{equation}{3}
For the simplification of the numerical procedure, we can rewrite the set of the
linear equations in the following way:
\begin{equation}
\sum^{n}_{k=1}A_{ik}x_{k}=B_{i}\Rightarrow\sum^{n}_{k=1}A^{'}_{ik}x_{k}=B^{'}_{i} \,.
\label{eq4}
\end{equation}
where we delineate, that:
\begin{equation}
A^{'}_{ik}=\frac{A_{ik}}{\sqrt{\sum^{n}_{k=1}A^{2}_{ik}}}\equiv A_{ik},
\quad\quad B^{'}_{i}=\frac{B_{i}}{\sqrt{\sum^{n}_{k=1}A^{2}_{ik}}}\equiv B_{i}\,.
\label{eq5}
\end{equation}
Here we want to underline, that within the normalization procedure, the $i$ index is fixed.
Thus, we can now rewrite the set of the linear equations using the same form, but now our coefficients are normalized
which means that for any indices pair of $i$, and $k$ we observe: $|A_{ik}|\leq 1$,
which leads to the conclusion that all of our vectors which are perpendicular to their hyperplanes are interrelated.

\section{A first arbitrary point}

For the iteration procedure we need to choose the first point,
from which the iteration procedure will start. It is obvious,
that the iteration procedure convergence will be dependent on this starting point.
It would be preferred to derive this point as close as possible to the final
solution. Which of course is impossible to do in a general case.  However, at a minimum we can require
that within the iteration procedure itself that the chosen initial starting point must be numerically stable.
In this case, we can start from a point which for example might belong
to the first hyperplane and have coordinates:$(x_1,0,0,...,0)$, where $x_1=\frac{B_1}{A_{11}}$.

\section{An iteration procedure}
\setcounter{equation}{5}
For the start of the iteration procedure we need in that procedure a method for the projection of any
arbitrary point $A$ (defined by vector $\vec{x}^{(1)}$) on the $i$ -th hyperplane. In
this case it is easy to see, that if the projection of this point is some point
 $P$ (defined by vector $\vec{x}^{(2)}$), which belongs to the $i$-th hyperplane, then:
\begin{equation}
\vec{x}^{(2)}=\vec{x}^{(1)}+\lambda\vec{A_i}\,.
\label{eq6}
\end{equation}
here $\vec{A}_i$ is a unit vector, which is $\perp$ to the $i$-th hyperplane.
Because $P\in(\alpha_i)$, then we can state that:
\begin{equation}
\sum^{n}_{k=1}A_{ik}(x^{(1)}_k +\lambda A_{ik})=
\lambda +\sum^{n}_{k=1}A_{ik}x^{(1)}_k=B_i
\label{eq7}
\end{equation}
From where we determine:
\begin{equation}
\lambda =B_i - \sum^{n}_{k=1}A_{ik}x^{(1)}_k
\label{eq8}
\end{equation}
and it follows that
\begin{equation}
|\vec{A}_i|=1\to\sum^{n}_{k=1}A_{ik}A_{ik}=1
\label{eq9}
\end{equation}
After determination of the projection parameter $\lambda$ we can now describe the components
of the projection point $P$:
\begin{equation}
x^{(2)}_k=x^{(1)}_k+\lambda A_{ik}
\label{eq10}
\end{equation}
where our iteration procedure consists of the following steps:\\
%\begin{romanlist}[(iii)]
%\item[(i)] 

1. The first step starts from the procedure of choosing the
1-st point, which will belong to the 1-st hyperplane and we will project it
on to the next or second hyperplane.\\
%\item[(ii)] 

2. The second step consists of the projection of the 2-nd point which belongs to the
2-nd hyperplane, on to the next or 3-rd hyperplane and so forth until the last
$n$ hyperplane is reached.\\
%\item[(iii)] 

3. The third step then consists of the projection of the iteration point from the
$n$-th hyperplane onto the first hyperplane, in other words the one which began our iteration procedure.
After that we then can repeat all of the above listed procedures again until a complete convergence
to the final solution is reached.
%\end{romanlist}

\section{Convergence}
\setcounter{equation}{10}

Proving the convergence in the numerical procedure described above can be accomplished in the following way.
If we take some arbitrary initial
point $A$, which belongs to a hyperplane and its projection point $P$
on the $i$-th hyperplane, then it is clear, that the distance between projection point $P$
and our final point of solution, noted as $S$, is: $L_{PS}=L_{AS}\cos\varphi$ therefore
for the $k$-th number of iterations we can to write, that: $L_{k}=L_{k-1}\cos\varphi_k$.
Then the distance between solution $S$ and our projection point $P$ after $n$ projections
can be written as:
\begin{equation}
L_{PS}=\lim_{n\to\infty} L_0(\cos\varphi_1\cos\varphi_2...\cos\varphi_n)\to 0
\label{eq11}
\end{equation}
For example, in the case, when all our hyperplanes are mutually orthogonal, then
we have, for any number $k$ a value of $\cos\varphi_k = 0$ and then we can automatically
obtain a solution after, at least for $n-1$  iterations. However, in a case in which
the difference between vectors, which happen to be $\perp$ to the hyperplanes is
$\Delta\varphi_{i,k} \to 0$ for any two given hyperplanes then in this case the number
of iteration procedures becomes $n\to\infty$. It is clear, from a geometric point
of view that in this case we have almost parallel hyperplanes and the convergence
will proceed slowly. It is possible however, to roughly estimate the approximate number of iterations:
\begin{equation}
 L= L_0<\cos\varphi >^n \approx \varepsilon
\label{eq12}
\end{equation}
from which we arrive at:
\begin{equation}
<\cos\varphi >= \frac{2}{\pi}\int^{\frac{\pi}{2}}_{0}\cos\varphi d\varphi =\frac{2}{\pi}\to
n\approx\frac{\ln\biggl(\frac{\varepsilon}{L_0}\biggr)}{\ln\biggl(\frac{2}{\pi}\biggr)}=
\log_{\frac{2}{\pi}}\biggl(\frac{\varepsilon}{L_0}\biggr)
\label{eq13}
\end{equation}
here $\varepsilon$  - is the accuracy of the solution. This example simply shows how
the number of iterations depends on the choice of the first point selected for any arbitrary
system, in general. However, the main result is that this iteration procedure is
numerically stable, it is not dependent on the choice of the initial starting point
to achieve convergence, if a solution to the set of linear systems exists.

\section{Numerical example}
\setcounter{equation}{13}

Let's consider a simple numerical example. We will use our method to
solve the following set of the linear equations:
\begin{eqnarray}
x_1+2x_2+3x_3+4x_4+5x_5=55\nonumber \\
-2x_1+10x_2+x_3+3x_4+8x_5=73\nonumber \\
4x_1+5x_2+8x_3+x_4+12x_5=102\nonumber \\
x_1+2x_2+x_3+3x_4+10x_5=70\nonumber \\
8x_1+17x_2+x_3+4x_4+3x_5=76
\label{eq14}
\end{eqnarray}
The exact solution for this set of the linear equations is:
\begin{eqnarray}
x_1=1;\quad x_2=2;\quad x_3=3;\quad x_4=4;\quad x_5=5
\label{eq15}
\end{eqnarray}
For the first starting point we have:
\begin{eqnarray}
x_1=55;\quad x_2=0;\quad x_3=0;\quad x_4=0;\quad x_5=0
\label{eq16}
\end{eqnarray}
The result of this iteration procedure is presented in Table 1. Based on these
results we are able to state that $\Delta=max|\sum^{n}_{k=1}A_{ik}x_{k}-B_i|$
is the maximum deviation from the solution at the given number of iterations
when $i\in \{1,2,3,4,5\}$.

%\begin{table}
%\caption{\label{tab:table2}Comparison of the energy of neutron's bound states (in {\it eV} units)
%on the NS and Earth's surface}
%\begin{ruledtabular}
%\begin{tabular}{ccc}

\begin{table}  %[ht]
%\tbl
\caption{Demonstration of the numerical procedure convergence.}
{\begin{tabular}{ccccccc}
\toprule
 && &Number of iterations& \\ \hline
&10     & 30         & 50         & 70          &100        &200\\ \hline
$x_1$   &6.36729084  & 0.24225408 & 1.00692936  &1.00648839 &0.99979049           & 1.0 \\
$x_2$   &-1.70446795 & 2.26483449 & 2.011421866 &1.99653770 &2.00010026           & 2.0 \\
$x_3$   &-5.01380157 & 3.74870823 & 3.02248973  &2.99148333 &3.00025466           & 3.0 \\
$x_4$   &13.2914602  & 4.28298482 & 3.89357617  &4.00503095 &3.99990957           & 4.0\\
$x_5$   &1.96186307  & 4.89304449 & 5.03535119  &4.99844825 &5.00002627           & 5.0\\ \hline
$\Delta$&5.9         & 0.3677     & 0.0377      &0.0046     &$7.073\cdot 10^{-5}$ & $1.178\cdot 10^{-9}$\\ \botrule
\end{tabular}}
\end{table}

\section{Conclusion}

In the current mathematical literature there exists many different and useful methods for obtaining a
solution for
a set of the linear algebraic equations. The most notable include works by Cramer, Gauss, Jacobi, Gauss - Seidel.
These methods provide a very useful solution to the problems described in the works \cite{Ilyin,Maksi,Volkov}.
However, a major problem arises in attempts to solve problems involving a very high number of variables,
for example when the number of variables is equal to $10^{4}-10^{6}$ or more. In this case we
need to allow more time for convergence, especially if we use an iteration method. However, at the same
time many of these existing methods become numerically unstable. For example, even a
solution of obtaining the inverse matrix to $A_{ik}$ becomes a non trivial problem.

The application of our method is rather simple at the programming level and it has a transparent
geometrical explanation and easy to understand pictorial representation. The efficiency of this method is much higher in
solutions involving problems with large numbers of variables. The iteration procedure works rather quickly, for
example in our case when the number of iterations is equal to $10^6$ it takes less then 5 sec.
In other methods the number of iterations required are sometimes much less than in our method, but they may
require some special properties and/or conditions.
%as it is in the $A_{ik}$ matrix. 
This is the case of the Jacobi method. Our approach is not sensitive to these problems.

%\section*{Acknowledgments}

\begin{acknowledgments}
A.K. express his gratitude to the Bogoliubov Laboratory of
Theoretical Physics (JINR, Dubna, Russia) for the invitation and warm hospitality.
A grant from OSP, St. Cloud State University (St. Cloud, Minnesota, USA) is also
gratefully acknowledged.
\end{acknowledgments}

\end{document}